\documentclass[reprint, amsmath,amssymb, aps]{revtex4-1}

\usepackage{graphicx}
\usepackage{dcolumn}
\usepackage{bm}
\usepackage[english]{babel}
\usepackage[utf8]{inputenc}
\usepackage{enumitem}
\usepackage{hyperref}
\usepackage{cleveref}

\begin{document}
 
\abovedisplayskip=-2pt
\abovedisplayshortskip=-2pt
\belowdisplayskip=7pt
\belowdisplayshortskip=7pt

\title{Nonlinear Acoustics - Perturbation Theory and Webster's Equation}

\author{Rogério Jorge}
\affiliation{Departamento de Física, Instituto Superior Técnico,\\ Av. Rovisco Pais 1, 1049-001 Lisboa, Portugal}
\email{rogerio.jorge@ist.utl.pt}


\begin{abstract}
Webster’s horn equation (1919) offers a one-dimensional approximation for low-frequency sound waves along a rigid tube with a variable cross-sectional area S(x). It can be thought as a wave equation with a source term that takes into account the nonlinear geometry of the tube. In this document we derive this equation using a simplified fluid model of an ideal gas. By a simple change of variables, we convert it to a Schrödinger equation and use the well-known variational and perturbative methods to seek perturbative solutions. As an example, we apply these methods to the "Gabriel's Horn" geometry, deriving the first order corrections to the linear frequency. An algorithm to the harmonic modes in any order for a general horn geometry is derived.
\end{abstract}

\maketitle

\section{Introduction}

\par We study the propagation of a wave in a narrow but long, tubular domain of finite length whose cross-sections are circular and of varying area. In this case, the wave equation has a classical approximation depending on a single spatial variable in the long direction of the domain. This approximation is known as Webster’s equation (\ref{webster}). The geometry of the tube is represented by the area function $S(x)$ whose values are cross-sectional areas of the domain. We derive this result in section II.
\par As the name suggests, this equation was derived by Webster in 1919 \cite{webster} but, citing Edward Eisner (referring to P. A. Martin article in \cite{onwebster})  - "we see that there is little justification for this name. Daniel Bernoulli, Euler, and Lagrange all derived the equation and did most interesting work on its solution, more than 150 years before Webster."
\par In section III, the link between the Schrödinger's equation and \cref{webster} is shown, offering an effective Hamiltonian and a potential energy that can be thought as a perturbation to the "free" hamiltonian.
\par The key feature of this document is the analysis in section V, where we obtain the first order harmonic corrections using perturbation theory on a well known geometry - Gabriel's horn (discussed in section IV).
\par In section VI, an algorithm to obtain this frequency corrections in any order and geometry is provided. With this analysis, we can infer how much the instrument (in the wind or brass family) will be  out of tune only by its geometry.

\section{Physical Model}

\subsection{Extended Derivation}

\par From fluid mechanics, the material derivative $\frac{D m}{Dt}$ is given by Reynolds' transport theorem. It can be stated as

\begin{equation}
\frac{D m}{Dt}= \partial_t \int _V\rho dV + \int _S (\vec v . \vec n) \rho dS=0,
\label{reynolds}
\end{equation}

\noindent where $m$ is the mass of the gas inside the tube (which is constant), $\rho$ is its mass density, $V$ and $S$ are the volume and surfaces of integration along the tube and $\vec v$ is the velocity of the gas in across the surface of integration.

\par
\begin{figure}
\includegraphics[width=\linewidth]{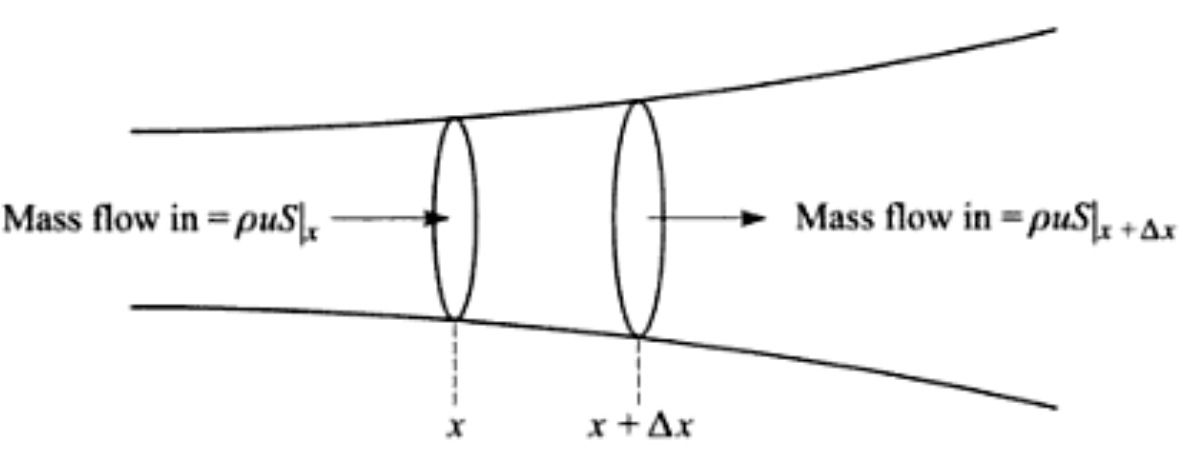}
\caption{Fluid model for a varying cross section. Image taken from \textit{Fundamentals of Physical Acoustics} by David T. Blackstock}
\label{horn}
\end{figure}

 Choosing these domains of integration and reference axes, we refer to \cref{horn}.

\par For a consistent and rigorous derivation, we assume the following conditions:

\begin{itemize}
\item{The mass density is constant throughout the cross-section area but is time dependent $\rho=\rho(t)$ (later we will simplify this assumption).}
\item{The tube shape is fixed i.e. independent of time but not constant in $x$, which by \cref{reynolds} implies $(\partial_t \rho) S=-\rho (\partial_x v S)$.}
\item{The ideal gas law $P= k_B T n$ holds, where $P$ is the pressure, $T$ the temperature (assumed constant), $k_B$ the Boltzmann constant and $n=\frac{\rho}{m}$, being $m$ the mass of each particle (assumed equal).}
\item{By Newton's second law - $\rho \frac{\partial v}{\partial t} = -\frac{\partial P}{\partial x}$.}
\item{The fluid is irrotational, meaning $\nabla \times \vec v=0$. Hence, differential calculus tells us that we can always find a velocity potential $\phi$, such that (in one dimension) $v =-\frac{\partial \phi}{\partial x}$.}
\end{itemize}

\par This offers a closed system of equations. Solving all equations for $P$, we discard at the end the time derivative of $\rho$. This is justified assuming that for long tubes, the local pressure variation is much larger than the local density fluctuation, obtaining Webster's equation (\cref{webster}).

\begin{equation}
\frac{1}{S}\frac{\partial}{\partial x}\left(S \frac{\partial P}{\partial x} \right)=\frac{1}{c^2}\frac{\partial^2P}{\partial t^2},
\label{webster}
\end{equation}

\noindent where $c^2=\frac{k_B T}{m}$. Physically, the measurable quantity in the laboratory is $P$, justifying the form of \cref{webster}.

\subsection{Alternative Derivation}

\par Using the concept of bulk modulus, we can easily derive (but lacking physical intuiton) equation (\ref{webster}). The differential volume for the gas section is $dV=dx \frac{\partial}{\partial x}\left(S \zeta\right)$, where $\zeta = \zeta(x,t)$ is the displacement of surfaces with equal pressure. By the ideal gas law, we can use the definition of bulk modulus $B$ (assumed constant) to obtain

\begin{equation}
P(x)=-\frac{B}{S}\partial_x \left(S \zeta\right).
\end{equation}

\par From Newton's second law we compute the gas volume acceleration due to pressure variation along $x$

\begin{equation}
S \rho dx \frac{\partial^2 \zeta}{\partial t^2} = - \frac{\partial P}{\partial z}S dx.
\end{equation}

\par Substituting into the previous equation we have Webster`s equation with $c^2= \frac{B}{\rho}$. This is also the procedure used in \cite{hornfunction}.

\section{Relation with Schrödinger Equation}

\par Starting with Webster's equation (\ref{webster}) previously derived, we apply the following change of variables (as in \cite{hornfunction})

\begin{itemize}
\item{$\psi = P \sqrt{S}$,}
\item{$r = \sqrt{S}$,}
\item{The time dependence on $P$ is given by $e^{i w t}$,}
\item{$k = \frac{w}{c}$,}
\end{itemize}

\noindent which in turn implies

\begin{equation}
-\frac{d^2 \psi}{dx^2}+\frac{r''}{r}\psi=k^2\psi.
\label{schrodinger}
\end{equation}

\par This is equivalent to the Schrödinger equation for one dimensional scattering, where the particle's energy is now $k^2$ and the potential energy function ise replaced by $\frac{r''}{r}$. In literature, this equation is often called the \textit{horn function}. The "potential energy" can be thought as a normalized curvature and $r$ is (apart from a numerical factor) the radius of the horn.

\par We can even infer a Hamiltonian operator from (\ref{schrodinger})

\begin{equation}
\hat{H}=\hat{T}+\hat{V}=-\frac{d^2}{dx^2} + \frac{r''(x)}{r(x)},
\label{hamiltonian}
\end{equation}

\noindent since $\psi$ has no time dependence. In our case, the tube is long (compared with its radius), so we can treat the potential $\frac{r''(x)}{r(x)}$ as a perturbation of the "free" hamiltonean.

\section{Gabriel's Horn}

\par Gabriel's horn, also called Torricelli's trumpet, is the surface of revolution of $y = \frac{1}{x}$ about the x-axis for $x \ge 1$. It is therefore given by the following parametric equations (\cite{wolfram})

\begin{equation}
x=u, \hspace{0.2cm} y=\frac{a \hspace{0.1cm} \cos(\nu)}{u}, \hspace{0.2cm} z=\frac{a \hspace{0.1cm}\sin(\nu)}{u},
\end{equation}

\noindent where $a$ is the radius of the surface on $x=1$. It is easy to show that this surface has finite volume, but infinite surface area.

\begin{figure}
\includegraphics[width=\linewidth]{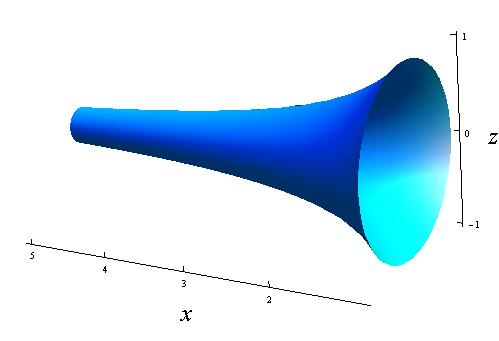}
\caption{Gabriel`s Horn with $a=1$ (computed with the \textit{Wolfram Alpha} platform).}
\label{gabriel}
\end{figure}

\par We can see that, for \cref{gabriel}, we have

\begin{equation}
r(x)=\frac{a}{x}, \hspace{0.2cm} S(x)=\frac{\pi a^2}{x^2},
\end{equation}

\noindent so Webster's and Horn equations (\ref{webster} and \ref{schrodinger}) become respectively (\ref{schorn}) and (\ref{gweb}).

\begin{equation}
\frac{d^2 P(x)}{dx^2} - \frac{2}{x}\frac{d P(x)}{dx} + w^2 P(x) = 0
\label{gweb}
\end{equation}

\vspace{-0.2cm}

\begin{equation}
\frac{d^2 \psi}{d x^2} + \left(k^2 - \frac{2}{x^2}\right)\psi=0.
\label{schorn}
\end{equation}

\par The solution for these equations are, respectively

\begin{equation}
P(x)= \sqrt{\frac{2}{\pi w^3}}\left[(A w x + B)\cos(w x) +(B w x- A)\sin(w x)\right]
\end{equation}

\begin{equation}
\psi(x) = \sqrt{\frac{2}{\pi k^3}}\left[\left(A k + \frac{B}{x}\right)\cos(k x) +\left(B k - \frac{A}{x}\right)\sin(k x)\right].
\end{equation}

\par Clearly, these can't be expressed as a Fourier sum. This could be achieved by expanding over a small parameter and we will do so in the next section. It is also hard to find the quantized frequencies without the use of a numerical method. As we will see, the perturbative method offers much more insight and simpler expressions to use on real situations.

\section{Perturbative Methods}

\par As stated in section III, \cref{hamiltonian} is the hamiltonian of the system, written as a sum of a "free" hamiltonian plus a perturbation. The latter term, for Gabriel's horn, is written as $\frac{2}{x^2}$.
\par The solution of \cref{schorn} for the free wave (neglecting the potential) is $\psi_0 = A \cos(k x) + B \sin(k x)$. From the boundary conditions $k$ will be quantized, so a sum over $k_n$ is performed, obtaining a Fourier decomposition as we would expect.

\begin{equation}
\psi_0 = \sum_n A_n \cos(k_n x) + B_n \sin(k_n x).
\end{equation}

\par The tube is open on both sides, which means, the pressure must be $0$ on $x=1$ and $x=L$ (defined as the constant ambient pressure). The origin is chosen so that the tube length is $L-1$ and the potential is regular.

\par At zeroth order, the surface is constant. Reverting the change of variables and denoting $S_0$ as the surface area at the origin, we have $P_0(x)=\frac{1}{\sqrt{S_0}}\sum_n (A_n \cos(k_n x) + B_n \sin(k_n x))$.

\par The boundary conditions impose the quantization $\tan(k)=\tan(k L)$ and the form (\ref{psio})

\begin{equation}
\psi_0=\sum_{n} B_n \left( \sin( k_n x) - \tan(k) \cos(k_n x) \right), \hspace{0.2 cm} k_n=\frac{n \pi}{L-1},
\label{psio}
\end{equation}

\noindent which in turn offers the expected result $f_n=\frac{c}{2 (L-1)} n$, where $f$ is the frequency of the sound wave and $n$ is an integer $>0$ and $L-1$ is the real length of the tube.\\

\par In the quantum mechanics formalism (as the one outlined below), the wave function must be normalized so that an explicit expression for $B_n$ can be found. This is the merging point of the classical and quantum treatment so it must be carefully done. This can be done by the following algorithm:

\begin{itemize}
\item{Obtain the pressure profile and the length $L-1$ of the horn. With this, compute $\lambda^2=\int_L P^2(x)dx$. Normalize the pressure profile by $\lambda$.}
\item{Performing the previous integral analitically, the factor $B_n$ will depend on $\lambda$. By the argument above, we can set $\lambda=1$ in our model.}
\end{itemize}

\par Normalizing the square of the wave function over the tube results in

\begin{equation}
B_n = \sqrt{\frac{2}{L-1}}\cos\left(\frac{n \pi}{L-1}\right).
\label{bn}
\end{equation}

\subsection{Variational Method}

\par
\begin{figure}
\includegraphics[width=\linewidth]{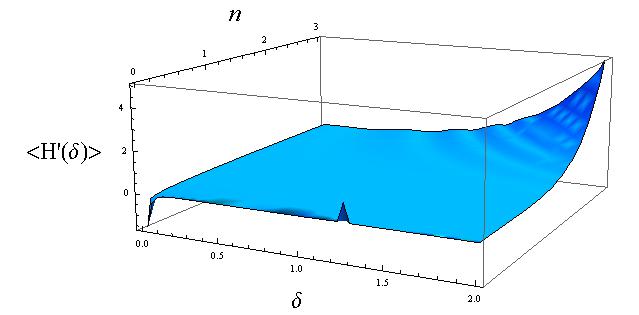}
\caption{$\frac{d}{d \delta}<H(\delta)>$ for a Gabriel horn with $L=20$ u.l., $\delta \in [0,2]$ and $n \in [0,3]$.} 
\label{variational}
\end{figure}

 This model requires a test function and a minimizing parameter $\delta$. As expected, our test function will be (\ref{psio}) (the free wave). The parameter, as defined in \cref{delta}) is expected to minimize $<H>$ near $\delta=1$. The function $\frac{d}{d\delta}<H>$ is shown in \cref{variational}. The results are valid for all $n$, as \cref{variational} implies (for bigger $n$ the derivative explodes).
\par For $\delta > 0$, there are no roots of $\frac{d}{d\delta}<H>$, so the method can't be applied in this framework. 
\par Incidentally, the variational method only provides a correction to the "ground state", so we can't calculate to an arbitrary order the corrections to the frequency.

\begin{equation}
k_n(\delta)= \frac{ \pi}{L-1}n^\delta
\label{delta}
\end{equation}

\subsection{Non-degenerate time independendent perturbation theory}

\par Perturbation theory tells us that the difference in energy $k^2$ from $k_0^2$ in first order is

\begin{equation}
\Delta k^2 =  \int_1^{L} \psi_0^\dagger \hat{H}\psi_0 dx = \int_1^{L}  \frac{2 \psi_0^2(x)}{x^2} dx,
\label{intpt}
\end{equation}

\noindent where $\psi_0$ is the unperturbed wave function.

\par
\begin{figure}
\includegraphics[width=\linewidth]{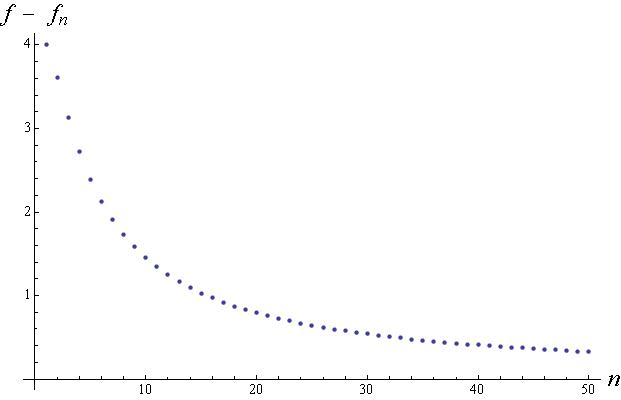}
\caption{Difference of the total frequency to the unperturbed one up to $n=50$ for a $L=20$ Gabriel horn, with $c=344$ m/s in first order perturbation theory.}
\label{ftotal}
\end{figure}

 Performing the integration on (\ref{integralbacano}), an analytical expression for $\Delta k^2=k^2-k_n^2$ is obtained. A plot of $f-f_n$ is shown on \cref{ftotal}.

\begin{equation}
\Delta k^2 = \frac{4 \cos^2\left(\frac{n \pi}{L-1}\right)}{L-1} \int_1^L \frac{\left[\sin(k x) - \tan(k) \cos(k x)\right]^2}{x^2}.
\label{integralbacano}
\end{equation}

\section{Concluding Remarks}

\par We have derived the expression for the perturbation on the frequency spectrum of a horn with varying cross section using time-independent perturbation theory in first order. Physically, the wave is an infinite sum of $n$ modes. Analyzing our results, the perturbation convergence is secured, as the correction is smaller for smaller values of $n$.

\par As a general procedure, one could find how much the geometry of the horn "constraints" the non-linearity of the frequency harmonics.

\begin{itemize}
\item{Express the radius of the horn in terms of the cylindrical coordinate $x$ - $r=r(x)$ and the length of the horn as $L-1$.}
\item{On that particular horn, measure the value of $\lambda^2 = \int_L P^2(x) dx$ and normalize the pressure profile by $\lambda$.}
\item{Using $B_n$ as in \cref{bn} with $\lambda=1$, $\psi_0 = B_n \left[\sin(k_0 x) - \tan(k_0) \cos(k_0 x)\right]$ and $k_0=\frac{n \pi}{L-1}$, calculate the integral $\Delta{k^2}=\int_1^L \psi_0^2(x) \frac{r''(x)}{r(x)}dx$.}
\item{The correction to the wave number in first order is $k=\sqrt{k_0^2+\Delta{k^2}}$.}
\end{itemize}

\par We hope that this approach serves both the acoustical science community and the curious physicist, providing an interesting application to the quantum mechanical methods within a classical framework.\\

\begin{acknowledgments}
I would like to thank prof. Henrique Oliveira for fruitful discussions and prof. Filipe Joaquim for essential corrections to the text and guidance. 
\end{acknowledgments}


\begin{thebibliography}{1}
\bibitem{wolfram}
Weisstein, Eric W. "Gabriel's Horn." From MathWorld--A Wolfram Web Resource. http://mathworld.wolfram.com/GabrielsHorn.html

\bibitem{webster}
Webster, A. G. (1919) ‘‘Acoustical impedance, and the theory of horns and
of the phonograph’’, Proc. Natl. Acad. Sci. U.S.A. 5, 275–282.

\bibitem{miksis}
Ting, L., and Miksis, M. J. (1983) ‘‘Wave propagation through a slender
curved tube’’, J. Acoust. Soc. Am. 74, 631–639.

\bibitem{onwebster}
P. A. Martin, (2004) "On Webster’s horn equation and some generalizations", Acoustical Society of America. [DOI: 10.1121/1.1775272]

\bibitem{audioxpress}
Bjørn Kolbrek, (2008) "Horn Theory: An Introduction, Part 1", article prepared for www.audioxpress.com

\bibitem{hornfunction}
David Berners and Julius O. Smith III (1994) "On the use of Schrödinger's equation in the analytic determination of horn reflectance", ICMC Preceedings in Sound Synthesis Techniques

\bibitem{griffiths}
D. J. Griffiths (2005) "Introduction to Quantum Mechanics", 2nd Ed. Prentice Hall 

\end{thebibliography}
\end{document}